\documentstyle[multicol,prl,aps,epsfig,psfig]{revtex}
\newcommand{\prL}{Phys.\ Rev.\ Lett.\ }
\newcommand{\pr}{Phys.\ Rev.\ }
\newcommand{\jpb}{J.\ Phys.\ B\ }
\newcommand{\phr}{Phys.\ Rep.\ }
\newcommand{\anp}{Ann.\ Phys.\ }
\newcommand{\jpa}{J.\ Phys.\ A\ }
\def\bea{\begin{eqnarray}}
\def\eea{\end{eqnarray}}
\def\be{\begin{equation}}
\def\ee{\end{equation}}

\begin{document}
\draft
\tighten
\author{Krzysztof Sacha$^{1,2}$ and
Jakub Zakrzewski$^2$}
\address{
$^1$Fachbereich Physik, Universit\"at Marburg \\
Renthof 6, D-35032 Marburg, Germany\\
 $^2$Instytut Fizyki imienia Mariana Smoluchowskiego, Uniwersytet
Jagiello\'nski,\\
 Reymonta 4, PL-30-059 Krak\'ow, Poland
}
\title{Driven Rydberg atoms reveal quartic level repulsion
}
\date{\today}

\maketitle

\begin{abstract}
The dynamics of Rydberg states of a hydrogen atom subject simultaneously 
to uniform static electric field and two microwave fields with commensurate 
frequencies is considered in the range of small fields amplitudes. In the
certain range of the parameters of the system the classical secular motion 
of the electronic ellipse reveals chaotic behavior. Quantum mechanically, 
when the fine structure of the atom is taken into account, the energy level
statistics  obey predictions appropriate for  the symplectic Gaussian 
random matrix  ensemble.

\end{abstract}
\pacs{PACS: 05.45.Mt, 32.80.Rm, 42.50.Hz}

\begin{multicols}{2}
Quantum chaos considers correlations between the quantal properties
of dynamical system and its classical behavior. In particular 
level statistics of quantum systems chaotic in the classical limit should
generically obey Random Matrix Theory (RMT) predictions \cite{mehta91}
according to the famous conjecture of
 Bohigas, Giannoni and Schmit
\cite{bohigas84}. The link between RMT and chaotic systems has been
quite fruitful for     quantum chaos studies
(see for reviews \cite{haake90,bohigas91}).

Depending on the symmetries of a given strongly chaotic system, statistical
properties of its quantum spectrum fall into one of the three classes known
from RMT: orthogonal, unitary and symplectic. The orthogonal class
 is typically associated with Hamiltonians invariant with respect to
 some generalized time-reversal symmetry (referred to also as an antiunitary
 symmetry), the corresponding statistical ensemble
 of random matrices is referred to as Gaussian Orthogonal Ensemble (GOE).
 In the absence of any such a symmetry the corresponding class is known
 as Gaussian Unitary Ensemble (GUE). Finally energy levels of 
half-integer spin systems with no geometrical symmetries and obeying the 
generalized time-reversal symmetry $T$ squaring to minus unity
($T^2=-1$) reveal twofold Kramers degeneracy and their 
statistical properties pertain to Gaussian Symplectic Ensemble (GSE). 

 The conjecture has been tested on a number of theoretical models
 (we refer the reader to reviews \cite{haake90,bohigas91} rather than
 numerous original papers).  Experiments in atomic systems 
 have been restricted
  to the orthogonal universality class \cite{zimmermann88} only.
  To break antiunitary symmetries in atomic species one needs, e.g., 
   a static magnetic field, 
 nonhomogeneous on the atomic
scale \cite{haake90} which is quite hard to realize.
For that reason
GUE type statistics have been observed experimentally for microwave
billiards (so-called {\it wave chaos} experiments)  
where the time-reversal symmetry could be
broken by applying some ferrite devises \cite{so95,stoffregen95}.

As shown by us recently \cite{sacha99b,sacha00b} atomic systems
also allow to generate GUE type statistics provided one uses not only
the static fields but also microwaves. Then, instead of
considering properties of eigenvalues of a given Hamiltonian $H$ (which is
not possible for time-dependent microwave perturbation) one considers
quasienergies of the Floquet operator ${\cal H}=H-i\hbar\partial/\partial t$
well defined for a periodic driving \cite{shirley65}. For appropriately
chosen combination
of elliptically polarized microwaves and a static electric field,
the (generalized) time-reversal symmetries are broken and GUE type
statistics may be observed. Even for weak fields when Floquet states
may be thought of as the perturbed principal quantum number $n_0$ 
hydrogenic manifold,
the typical splitting between levels may be of the order of a few MHz
making the experiment feasible. A RMT type level statistics is a
manifestation of the classically chaotic behavior as observed (via classical
perturbation theory) for the motion of the electronic ellipse (the so-called
secular motion).

With GOE or GUE type of statistics realized for atomic species it seems
 natural to ask whether it is possible to observe also statistics
corresponding to the third universality class -- the symplectic ensemble.
The aim of this letter is to provide an example of such a situation.
 Such a behavior has not yet been observed, as far as
we know, neither for atomic nor for billiard type 
systems.

To observe the effect
one needs an interaction explicitly depending on the electron
half-integer spin, an interaction sufficiently intense to relatively
strongly perturb the levels. For hydrogen atoms
two obvious candidates are the Zeeman
effect and the spin-orbit coupling. 
In the former case, however, there is no time-reversal symmetry squaring
to minus unity.
The latter seems quite weak in the semiclassical Rydberg states regime.
For that reason traditionally the spin-orbit coupling is neglected while
considering strongly externally perturbed Rydberg states.

It is another story if we restrict ourselves to small external perturbations
and consider the combined effect of these perturbations and the spin-orbit
coupling. Our previous experience \cite{sacha99b,sacha00b} still
indicates that the secular motion of the electronic ellipse may be strongly
chaotic even for weak (but judiciously chosen) combination of the external
fields. Then, if also spin-orbit coupling is taken into account, one may hope
to observe GSE type statistics.

Thus we consider the very same model as before, i.e.,
the hydrogen Rydberg atom driven by microwaves and placed in a
static uniform electric field and we add the spin-orbit interaction.
The configuration of the external fields must be chosen so that: the
time-reversal symmetry with $T^2=-1$ is preserved, any other geometrical
symmetry is broken and the dynamics is irregular classically. 
To fulfill the first condition we have to use microwaves of
linear polarization, in contrast to our earlier 
studies \cite{sacha99b,sacha00b}. 
The Hamiltonian of the system in atomic units is
\be
H=\frac{{\mathbf p}^2}{2}-\frac{1}{r}+H_1+\alpha^2H_2,
\label{ham}
\ee
where
$\alpha$ is the fine structure constant.
$H_1$ is the external perturbation
\be
H_1={\mathbf E}\cdot {\mathbf r}+Fx\cos\omega t+F' z \cos 2 \omega t
\label{h1}
\ee
while
\be
H_2=-\frac{{\mathbf p}^4}{8}+\frac{1}{2r^3}
{\mathbf L}\cdot {\mathbf S}+\frac{\pi}{2}\delta({\mathbf r}),
\label{h2}
\ee
gives the lowest order relativistic corrections \cite{cct97}.
 Among them the most important
for the present contribution is the spin-orbit coupling. In the above formulae
${\mathbf E}$ denotes a static electric field vector, $F$ and $F'$ are
amplitudes of two
microwave fields with frequencies $\omega$ and $2\omega$
and  polarized along the $Ox$ and $Oz$ axes, respectively.

The Hamiltonian (\ref{ham}) is time reversal-invariant, with time-reversal
operator squaring to minus unity. Thus eigenvalues of the Floquet Hamiltonian
must reveal a twofold Kramers degeneracy \cite{haake90}.
The role of $F'$ field is to break the geometrical symmetry which is 
preserved by
(\ref{ham}) if $F'=0$. Indeed if we choose, for $F'=0$, the $Oz$ axis such that 
a static field component parallel to it vanishes, $E_z=0$, then the parity
operator
$\Pi_z=\exp{(i\pi S_z)}P_z$ (where  $P_z$ is a reflection with respect to
$Oxy$ plane) with the property $\Pi_z^2=-1$ 
commutes with $H$. This unitary symmetry splits the quasienergy
spectrum into two blocks with identical eigenvalues (Kramers degeneracy).
Each block is in itself a hermitian matrix, i.e. its eigenvalues are 
expected to obey GUE but not GSE statistics if the underlying classical 
motion is chaotic \cite{haake90}.

Since both the microwaves and the static electric field are assumed to be
weak, the model may be analyzed perturbatively both classically and
quantum mechanically. The latter is straightforward in the effective
Hamiltonian formalism \cite{cct92} with the fields taken in the 
lowest nonvanishing
order (first order for the static electric field due to a linear Stark
effect as well as for the spin-orbit interaction
 and second order in microwave amplitudes $F$ and $F'$). The resulting matrices
of dimension $2n_0^2$ with $n_0$ being the principal quantum number
of the perturbed manifold studied have been diagonalized with standard
routines. We refer the reader to \cite{sacha99b} for practical details
  described there for a similar model.

\begin{figure}
\centering
\epsfig{file=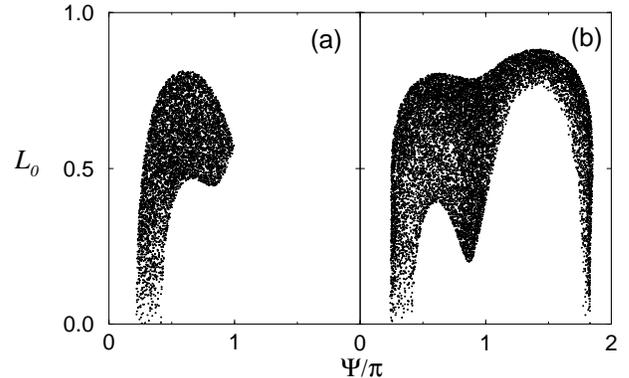,scale=0.34,angle=-90}
\caption{Poincar\'e surface of section (at $\Phi=0$) 
of the classical secular motion, Eq.~(\protect{\ref{final}}), for the 
hydrogen atom in static electric and microwave fields for the energy
$(H_0^{\rm eff}+1/2)/E_0=-1$ and with the amplitude and frequency of the 
microwave field
$F_{0}^2/E_0=2$, $\omega_0=1.304$, respectively.
The coordinates used for the plot are the scaled angular momentum $L_0=L/n_0$ 
and its canonically conjugate angle $\Psi$.
Panel (a) corresponds to the orientation of the static electric
field vector $\theta=\pi/2$, $\varphi=\pi/4$ while panel (b) is related to the
system additionally perturbed by $F'$ field
with the amplitude $F'^2_0/E_0=5$ (the static field orientation is 
$\theta=0.31\pi$, $\varphi=\pi/4$).
Note that, for the parameters chosen,
not the whole $(L_0,\Psi)$ space is accessible.
}
\label{one}
\end{figure}

In an analogous way one may construct an effective classical
Hamiltonian averaging over the phase of the field and 
the fast motion along the electronic ellipse 
and considering the slow secular motion of the ellipse
itself only. For the classical analysis we neglect the relativistic
correction part $H_2$ (the spin-orbit interaction and Darwin term,
$\delta({\mathbf r})$,
do not have direct classical analogs). For the
purpose of the present analysis we assume the microwaves to be off
resonant thus the perturbation calculations follow the Lie approach
 of \cite{lichtenberg83,abu97,sacha00b} rather than
that appropriate for resonant driving \cite{sacha99b}.

 The resulting classical effective Hamiltonian (first order in $E$ and second
 order in $F$ and $F'$) takes a form
\be
H^{\rm eff}_0=
-\frac{1}{2}+H_{1,0}^{\rm eff}(\omega_0,F_0^2,F'^2_0,E_0;L_0,\Psi,M_0,\Phi)
\label{final}
\ee
where $H^{\rm eff}_0=n_0^2H^{\rm eff}$, $F_0=n_0^4F$, $F'_0=n_0^4F'$, 
$E_0=n_0^4E$, $\omega_0=n_0^3\omega$, $L_0=L/n_0$ and
$M_0=M/n_0$  are scaled variables ($L,\ M$ being the angular momentum
 and its projection on the
$Oz$ axis while $\Psi$ and $\Phi$ are angles conjugate to $L$ and $M$,
respectively). Observe that classical dynamics 
depends only on the reduced energy via scaled variables 
(no dependence on $n_0$). Fig.~\ref{one}a shows an example of Poincar\'e
surface of section obtained for $(H^{\rm eff}_0+1/2)/E_0=-1$ with 
$F'_0=0$, $F_0^2/E_0=2$, $\omega_0=1.304$ and orientation of the static field
vector $\theta=\pi/2$, $\varphi=\pi/4$ (where $\theta,\ \varphi$
are usual spherical angles). In Fig.~\ref{one}b there is a similar plot but 
for the case with broken the parity symmetry, 
i.e. for the same parameters as previously but with  
$F'^2_0/E_0=5$ and $\theta=0.31\pi$. Clearly the motion is 
predominantly chaotic in both cases.

\begin{figure}
\centering
\epsfig{file=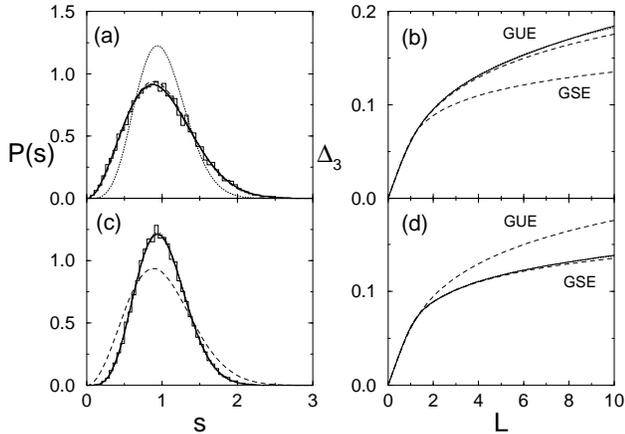,scale=0.34,angle=-90}
\caption{ Nearest neighbor spacing distribution and spectral rigidity,
$\Delta_3$ statistics, for hydrogen atom placed in a static electric field
and illuminated by microwave fields, for the case with [panels (a)-(b)]
and without [panels (c)-(d)] the parity symmetry $\Pi_z=\exp{(i\pi S_z)}P_z$
compared with the predictions of random matrices ensembles. 
In panels (a) and (c) solid lines indicate the best fitting 
Izrailev distributions, while dashed and dotted lines 
correspond to GUE and GSE distributions respectively
(in panel (c) the dotted line is hardly visible behind the solid one). 
Panels (b) and (d): solid and dotted (hardly
visible behind the solid lines) lines correspond to numerical 
data and their best fits, while dashed lines indicate GUE and
GSE predictions as indicated in the figure. 
}
\label{two}
\end{figure}

Let us now consider the influence of the spin-orbit interaction in
quantum case.
The fine structure splitting scales as $n_0^{-3}$. Thus, when collecting
eigenvalues for different $n_0$ manifolds (to improve the statistics)
external fields have to be appropriately rescaled 
for the data to correspond to the same interesting physical situation
in which external perturbations are comparable to the spin-orbit
interaction. In effect we have diagonalized the quantum effective Hamiltonian
for different hydrogenic manifolds in the range $n_0=50-59$ 
keeping constant $n_0E_0=10^{-6}$, $F_0^2/E_0$ and $F'^2_0/E_0$ what 
ensures that the energy shift, with respect to the $-1/2n_0^2$ value, 
multiplied by $n_0^3$ is of the same magnitude for each $n_0$ manifold. 
For the case when $H$ is invariant with respect to the parity 
symmetry, $\Pi_z$ we have collected single members from each Kramers pair 
in the range  of ``$n_0^3$ $\times$ energy shift with respect to 
$-1/2n_0^2$'' between $-2\cdot 10^{-6}$ and $3\cdot10^{-7}$ while for 
the broken symmetry case between $-3\cdot 10^{-6}$ and $-10^{-6}$.

\begin{figure}
\centering
\epsfig{file=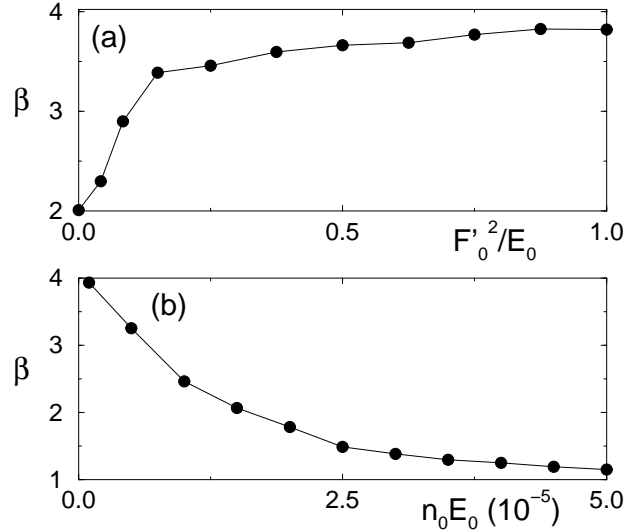,scale=0.44,angle=-90}
\caption{ The best fitting Izrailev distribution parameter, for hydrogen 
atom in the static electric field and driven by the two microwave fields, 
as a function of the parameters of the fields. Panel (a): gradual breaking of
the parity symmetry, $\Pi_z$ with a change of the amplitude $F'_0$ and 
for fixed: 
$F_0^2/E_0=2$, $n_0E_0=10^{-6}$, $\theta=\pi/2$ and $\varphi=\pi/4$. For 
$F'_0=0$ the symmetry is preserved, when $F'_0$ increases the symmetry is 
gradually broken and the statistics changes from the GUE to GSE type.
Panel (b): {\it switching off} the spin-orbit interaction -- keeping the
configuration of the external fields fixed (i.e. $F_0^2/E_0=2$, $F'^2_0/E_0=5$, 
$\theta=0.31\pi$ and $\varphi=\pi/4$) but increasing the fields amplitudes
the spin-orbit interaction becomes relatively weaker. This results in a change
of the level statistics from GSE to GOE type. To calculate each point in the
panels the effective Hamiltonian has been diagonalized for different hydrogenic
manifold in the range $n_0=55-59$.
}
\label{three}
\end{figure}

When the Hamiltonian is invariant with respect to the $\Pi_z$ transformation 
the nearest-neighbor spacing (NNS) distribution, $P(s)$, obtained is
fitted quite satisfactorily with the distribution appropriate for GUE
see Fig.~\ref{two} -- top row. The solid line is a best fitting Izrailev 
distribution \cite{izrailev90,casati91}
with the parameter $\beta=1.91$ 
($\beta$ gives the level repulsion exponent, 
i.e., $P(s)\propto s^\beta$ for small spacing $s$)
and with $\chi^2/N=0.4$, i.e.
chi-squared divided by the number of levels $N$ (there are about 14,000
levels in the data set). The spectral rigidities, $\Delta_3$,
also reveals the behavior very close to GUE -- the best fitting $\Delta_3$ 
statistics corresponding to an independent superposition of Poisson and GUE
spectra \cite{bohigas84} results in the parameter value (a relative 
measure of the chaotic part of the phase space) $q=0.99$.

Breaking of the parity symmetry has a dramatic effect on the quasienergies 
statistics as depicted in Fig.~\ref{two} -- bottom row.
This time
both NNS distribution and $\Delta_3$ statistics obtained show a
very good agreement with the predictions corresponding to GSE. In
particular one obtains a quartic level repulsion, $P(s)\propto s^4$
for small spacings. More precisely we have got 
the fitted  Izrailev distribution
repulsion parameter $\beta=3.86$ with $\chi^2/N=0.6$ and $q=1$ from the best
fitting $\Delta_3$ statistics (now it corresponds 
to an independent superposition of Poisson and GSE spectra \cite{bohigas84}).
This time there are about 20,000 levels in the data set.

Finally we would like to investigate gradual breaking of the parity
symmetry $\Pi_z$ as well as switching off the spin-orbit 
interaction \cite{bohigas84}. 
To this end we have diagonalized the effective Hamiltonian 
changing the parameters of the system.
In Fig.~\ref{three}a we show how the best 
fitting Izrailev distribution repulsion 
parameter, $\beta$ changes when $F'_0$ increases. That 
corresponds to a gradual breaking of the parity invariance -- the transition 
from GUE type statistics to GSE one is clearly visible in the figure.
{\it Switching off} of the spin-orbit interaction has been realized by 
increasing the microwave amplitudes which implies that the influence of the 
external fields
becomes stronger while the spin-orbit interaction becomes relatively
weaker. The results, as shown in Fig.~\ref{three}b, indicate the transition 
from GSE, through intermediate, to GOE type statistics. For negligible 
spin-orbit interaction additional unitary symmetries are restored (e.g. 
$\Pi_i=\exp{(i\pi S_i)}$ where $i=x,y$ with the property $\Pi_i^2=-1$) 
which allows splitting the Floquet Hamiltonian into two identical real
matrices \cite{haake90,caurier} and consequently 
the GOE statistics is expected. Actually, without the spin-orbit
interaction, the spin degrees of freedom have no effect on a dynamics
and can be eliminated. Then it becomes immediately clear that, because of the
time-reversal invariance, the system should reveal the GOE type statistics. 

To summarize, we have given an example of a realistic physical
system which may yield a quartic level repulsion and, more generally,
have statistical properties close to the Gaussian Symplectic Ensemble
of RMT. This is possible by taking into account the spin-orbit
interaction while using external perturbations (microwave and static)
for creating a chaotic secular motion of the electronic
ellipse. For the effect to
be visible the spin-orbit and external perturbations have to be
of comparable importance. This requires {\it very weak} external
perturbation since the spin-orbit coupling decreases rapidly with $n_0$.
The resulting mean Floquet levels spacing within the $n_0$
manifold is very small -- on the edge of present experimental possibilities for
unambiguous spectroscopy. 
To improve experimental conditions one may employ hydrogen-like ions since
the spin-orbit interaction increases with the charge of the nucleus and 
consequently mean level spacing could be greater.
Importantly, the Rydberg states in question
have large spontaneous emission lifetimes so, at least in principle,
the individual lines should be observable (and quartic level
repulsion assures low probability of small spacings).

We thank Dominique Delande, Bruno Eckhardt,  and Imre Varga for 
discussions. Support of KBN under project 2P302B-00915 
is acknowledged. KS acknowledges support by the Alexander von 
Humboldt Foundation.


\end{multicols}

\end{document}